# Uniform Spaces in the Pregeometric Modeling of Quantum Non-Separability

*W.M. Stuckey, Dept of Physics*
*Michael Silberstein, Dept of Philosophy, Elizabethtown College*
*Elizabethtown, PA  17022*

## Abstract

We introduce a pregeometry employing uniform spaces over the denumerable set X of spacetime events. The discrete uniformity $D_X$ over X is used to obtain a pregeometric model of macroscopic spacetime neighborhoods. We then use a uniformity base generated by a topological group structure over X to provide a pregeometric model of microscopic spacetime neighborhoods. Accordingly, quantum non-separability as it pertains to non-locality is understood pregeometrically as a contrast between microscopic spacetime neighborhoods and macroscopic spacetime neighborhoods. A nexus between this pregeometry and conventional spacetime physics is implied per the metric induced by $D_X$. A metric over the topological group $Z2 \ x \ \ldots \ x \ Z2$ is so generated. Implications for quantum gravity are enumerated.

**Keywords:** Uniform spaces, pregeometry, quantum non-separability, quantum

        non-locality, quantum gravity

## 1. INTRODUCTION

Uniform spaces lie between topological spaces and metric spaces in structural sophistication. Unlike the open sets of topological spaces, one may compare sizes of the entourages of a uniform space, although the size of an entourage needn't be specified via a mapping to ú as with a metric space. In some sense, a metric space is a uniform space with 'choice of gauge'. Therefore, uniform spaces provide a reasonable mathematics for pregeometry (Bergliaffa *et al.*, 1998).

Herein, we employ the discrete uniformity $D_X$ over the denumerable set X of spacetime events as a discrete, pregeometric model of macroscopic spacetime neighborhoods, i.e., spacetime neighborhoods compatible with the locality implicit in the construct of trans-temporal objects (Stuckey, 2000). We also show that $D_X$ induces a metric over finite X which may facilitate the transition from pregeometry to conventional spacetime physics. A metric so generated over the topological group Z2 x … x Z2 is shown to exhibit order in concert with its group structure.

To produce a pregeometric basis for quantum mechanics (QM) we note that $D_X$ induces the discrete topology, so the addition of a group structure over X yields a topological group. This topological group over X generates a uniformity U with base $U_B$ which provides a pregeometric model of microscopic spacetime neighborhoods. Quantum non-separability (QNS), as it relates to spatio-temporal non-locality, is then understood pregeometrically as a contrast between microscopic spacetime neighborhoods and macroscopic spacetime neighborhoods. Thus, uniform spaces $D_X$ and U over the denumerable set X of spacetime events provide a mathematical structure for the pregeometric basis of M4 which is distinct from, but consistent with, that of QM.

This work is motivated in part by the belief that innovative models of QNS can intimate properties of quantum gravity. This is neither a new nor unique tenet. For example, Heller and Sasin's approach to quantum gravity leads them to write (1998), "The correlations of the EPR type should be regarded as remnants of the totally non-local physics below the Planck threshold which is modeled by a noncommutative geometry."

"Correlations of the EPR type" refers to correlated, space-like separated experimental outcomes which violate Bell's inequality. "EPR," since correlations of this type were introduced by Einstein, Podolsky, and Rosen (1935). Bell (1964) explained how EPR correlations are more than a metaphysical issue when he showed there are realizable consequences. And in the early 1980's, Aspect *et al.* (1981, 1982) published experimental results confirming the violation of Bell's inequality and validating the predictions of QM.

While QM has been experimentally vindicated, many do perceive a conceptual problem regarding our understanding of spacetime as a consequence of QNS. How is it that particles travel through space, yet remain "non-separated?" That is, How do particles 'conspire' to produce quantum results in space-like separated detection events? In this sense, QNS bears directly on the modeling of spacetime - a key issue for quantum gravity. Demaret *et al.* write (1997):

> In this Section we analyse fundamental concepts of quantum mechanics. We show that they lead to some problems which force us to modify the usual notion of space-time. ... The second problem is related to the famous E.P.R. paradox which introduces the idea of non-locality or more precisely of non-separability with respect to space. In fact, in quantum mechanics space cannot be viewed as a set of isolated points. These problems lead to a deep modification of our representation of "quantum" space-time.

With this, Demaret and Lambert join Heller, Sasin, and others in expecting quantum conundrums to be resolved by quantum gravity. Unfortunately, the characteristics of quantum gravity have not been easy to discern. Weinberg writes (1999), "How can we get the ideas we need to formulate a truly fundamental theory, when this theory is to describe a realm where all intuitions derived from life in space-time become inapplicable?" We suggest that since quantum gravity will likely resolve quantum conundrums, its properties might conversely be sought in novel models of QNS.

## 2. PREGEOMETRY VIA UNIFORM SPACES

Perhaps the most innovative alternatives to conventional quantum gravity programs are those of pregeometry (for a review of pregeometry programs see Gibbs, 1995). According to Wheeler (Misner *et al.*, 1972), the locally M4 spacetime manifold of general relativity with its geometric counterparts to dynamic entities need not serve at the foundation of a reductionist model of reality. Rather, properties of the spacetime manifold, such as continuity, dimensionality, topology, and causality, might evolve mathematically from modeling fundamental to that of classical spacetime dynamics. Wheeler named this radically reductionist approach to quantum gravity *pregeometry*. Pregeometry is fecund and well suited for addressing QNS, since it is *ipso facto* unadulterated by conventional notions of locality.

In choosing a mathematical formalism for pregeometry, it seems prudent to avoid mathematics steeped in locality, e.g., differential calculus, the differential manifold, and the field structure over the set of reals. In order to avoid such local structures, a paragon of pregeometry might start with an unordered set X of spacetime events and develop explicitly the relative 'placement' of its elements in a mathematically natural fashion. It might also be independent of the domain of discourse for the underlying set *a la* general covariance. The uniform space provides just such a structure.

A uniform space U can be produced via any topological group, and a topological group can be constructed over any group by choosing the discrete topology over the underlying set X. While it is not true in general that a uniform space will produce a metric space, a metric may be generated from $D_X$ over denumerable, finite X. Given the microscopic realm is quantum, discrete mathematics would seem a reasonable choice for quantum gravity. Per Butterfield and Isham (1999), "For these reasons, a good case can be made that a complete theory of quantum gravity may require a revision of quantum theory itself in a way that removes the *a priori* use of continuum numbers in its mathematical formalism." And, Au writes (1995), "One can see how a discrete theory could reduce to a continuum one in the large scale limit, but to shed light on a discrete theory while working from the perspective of a continuum one seems difficult to achieve." After enumerating reasons to consider discrete mathematics in the fundamental modeling of spacetime, Sorkin concluded (1995), "The dynamical principles learned from quantum mechanics just seem to be incompatible with the idea that gravity is described by a metric field on a continuous manifold."

So, we are guaranteed that a uniform space U may be constructed over any denumerable set X by introducing a group structure and the discrete topology over X. And, $D_X$ induces the discrete topology over X while its entourages provide a conventional, but non-metric, definition of a ball centered on x ∈ X. Thus, $D_X$ induces the topology required for U while providing a pregeometric definition of macroscopic spacetime neighborhoods. Given that the introduction of a group structure over X underlying $D_X$ provides a uniformity base $U_B$ for U, we have the means to define microscopic spacetime neighborhoods independently of, but consistently with, macroscopic spacetime neighborhoods. The contrast between microscopic and macroscopic spacetime neighborhoods provides a pregeometric model of QNS.

**3. THE MODEL**

Specifically (Engelking, 1989), for x and y elements of X, a symmetric entourage V is a subset of X x X such that for each $(x, y) \in V$, $(y, x)$ is also an element of V. $D_X$ is the collection of all symmetric entourages. For $(x, y) \in V$ *the distance between x and y is said to be less than V. The ball with center x and radius V is* $\{y \in X \mid (x, y) \in V\}$ and is denoted B(x,V). A neighborhood of x in the topology induced by $D_X$ is Int B(x,V), so all possible balls about each $x \in X$ are established. This is precisely in accord with the conventional notion of locality, i.e., open balls about elements of the spacetime manifold. Therefore, B(x,V) is a perfect pregeometric definition of a macroscopic spacetime neighborhood of x for denumerable X (cf. Sorkin's finitary topological spaces (1991)). And, this definition of macroscopic spacetime neighborhoods accommodates the topological priority of causal chains over metric balls per Finkelstein (1969).

Such priority is supported by the process in which $D_X$ induces a metric over denumerable, finite X. We borrow from a proof of the following theorem as quoted from Engelking (1989):

"For every sequence $V_0, V_1, \ldots$ of members of a uniformity on a set X, where

$$V_0 = X \times X \text{ and } (V_{i+1})^3 \subset V_i \text{ for i} = 1,2,\ldots,$$

there exists a pseudometric $\rho$ on the set X such that for every $i \geq 1$

$$\{(x, y) \mid \rho(x, y) < (\tfrac{1}{2})^i\} \subset V_i \subset \{(x, y) \mid \rho(x, y) \leq (\tfrac{1}{2})^i\}.\text{"}$$

To find $\rho(x, y)$, consider all sequences of elements of X beginning with x and ending with y. For each adjacent pair $(x_n, x_{n+1})$ in any given sequence, find the smallest member of $\{V_i\}$ containing that pair. [The smallest $V_i$ will have the largest i, since $(V_{i+1})^3 \subset V_i$.] Suppose $V_m$ is that smallest member and let the 'artificial' distance between $x_n$ and $x_{n+1}$ be $(\tfrac{1}{2})^m$. Summing for all adjacent pairs in a given sequence yields an 'artificial' distance between x and y for that particular sequence. According to the theorem, $\rho(x, y)$ is the greatest lowest bound obtained via the sequences. When applied over denumerable, finite X the greatest lower bound will be non-zero, so the result will be a metric.

Obviously, without a formalism satisfying full correspondence with general relativity and QM one is free to speculate on alternative pregeometries based on this uniform space. For example, a pregeometric basis for "classical stochastic dynamics" (Rideout and Sorkin, 2000) is suggested by the combinatorial nature of $\rho(x, y)$. From there one might introduce an algebraic structure to account for non-locality *a la* Raptis and Zapatrin (2000). However, such an approach assumes a pregeometric basis for the stochasticity of QM. While it seems unlikely that quantum gravity will provide a non-stochastic basis for QM, we want to allow for the possibility that the theory fundamental to QM may harbor non-probabilistic attributes. Thus, we examine another uniformity for the pregeometric basis of QNS.

As stated previously, the introduction of a group structure G over X underlying $D_X$ allows for the construct of U. This is accomplished by constructing a uniformity base $U_B$ of U via neighborhoods of the identity e of G in the following fashion (Geroch, 1985). The entourage $A_\beta$ of U is $\{(x, y) \in X \times X \mid xy^{-1} \in \beta\}$ where $\beta$ is a neighborhood of e in the topology over X. When X is denumerable of order N, $\{(w, y) \in X \times X \mid w \neq y\}$ is partitioned equally into the entourages $A_x$ ($x \in X$ such that $x \neq e$) for the N - 1, order-two neighborhoods of e, i.e., $A_x$ is generated by $\{e, x\}$. The entourages $A_x$ and $\Delta \equiv \{(x, x) \mid x \in X\}$ constitute a base $U_B$ for U. Entourages generated by larger neighborhoods of e are given by members of $U_B$, i.e., $\{e, x, y\}$ generates $A_x \cup A_y$, etc.

While for some groups all members of $U_B$ are elements of $D_X$, as with the Klein 4-group (Stuckey, 1999), this is not true in general. In fact, $A_x \in D_X \ \forall \ x \in X$ such that $x = x^{-1}$. This, since for $(y, z) \in A_x$ such that $y \neq z$, $yz^{-1} = x$ and therefore, $zy^{-1} = x^{-1} = x \Rightarrow (z, y) \in A_x$. For the base members $A_x$ and $A_y$ such that $x = y^{-1}$, we have $A_x^{-1} = A_y$ where $A^{-1} = \{(w, z) \mid (z, w) \in A\}$. This, since for $(w, z) \in A_x$ such that $w \neq z$, $wz^{-1} = x$ and therefore, $zw^{-1} = x^{-1} = y \Rightarrow (z, w) \in A_y$. We may now construct the largest element of $D_X$ via multiplication of the members of $U_B$.

With $\Delta$ a subset of any entourage (uniquely and axiomatically), we have in general for entourages $A$ and $B$ that $A \subset AB$ and $B \subset AB$ where $AB \equiv \{(x, z) \mid (x, y) \in A \text{ and } (y, z) \in B\}$. Next, consider $\{(x, y), (y, z) \mid (x, y) \in A_s \text{ and } (y, z) \in A_w \text{ with } x \neq y \text{ and } y \neq z\}$. In addition to $\Delta$, these account exhaustively for the elements of $A_s$ and $A_w$. For any such pair $(x, y)$ and $(y, z)$, $(x, z) \in A_s A_w$ by definition and $(x, z) \in A_{sw}$, since $sw = (xy^{-1})(yz^{-1}) = xz^{-1}$. The N pairs $(x, z)$ with $\Delta$ account exhaustively for the elements of $A_{sw}$ and, excepting the impact of $\Delta$ on $A_s A_w$, the N pairs $(x, z)$ account exhaustively for the elements of $A_s A_w$. Again, the impact of $\Delta$ on $A_s A_w$ is to render $A_s \subset A_s A_w$ and $A_w \subset A_s A_w$. Therefore, $A_s A_w = A_s \cup A_w \cup A_{sw}$.

So, if $G$ is cyclic with generator $x$, $A_x^{N-1} = A_x \cup A_y \cup ... \cup A_z$, where $y = x^2$ and $z = x^{N-1}$. $A_x \cup A_y \cup ... \cup A_z$ is of course the largest element $V_{max}$ of $D_X$. [This is of particular interest, since the cyclic group structure $Z_N$ exists for all $N \in \mathbb{u}$ and is the unique group structure for N prime.] If $G$ is not cyclic, one may produce $V_{max}$ via $A_x \cup ... \cup A_y \cup A_w A_z \cup ... \cup A_s A_v$ for $x = x^{-1}, ..., y = y^{-1}, w = z^{-1}, ..., s = v^{-1}$, since $A_w A_z = A_w \cup A_z$ when $w = z^{-1}$. We are also guaranteed to produce $V_{max}$ via some variation of $A_x A_y ... A_z$ where $\{x, y, ..., z\} = X$, according to $G$. It should also be noted that, as implied *supra*, the entourage $A_x \cup A_y \cup ... \cup A_z$ is generated by the entire set X.

Should we define microscopic spacetime neighborhoods with the members of $U_B$ analogously to macroscopic spacetime neighborhoods per the symmetric entourages of $D_X$, we note the following interesting consequences. First, $A_s = A_w^{-1}$ for $s = w^{-1}$, so when $s \neq w$ *the distance between elements of $A_s$ is non-separable from that of $A_w$*, lest we compromise the symmetry of our pregeometric notion of distance. For $s = w$, i.e., $s = s^{-1}$, $A_s \in D_X$ and our microscopic spacetime structure accommodates separability. Thus, the degree to which our spacetime is to accommodate QNS is determined by the choice of G. Second, the choice of G over X underlying $D_X$ is all that is needed to produce the microscopic spacetime structure embedded in the macroscopic spacetime structure. Third, the members of the base $U_B$ of the microscopic spacetime structure U may be combined via entourage multiplication to yield the largest element $V_{max}$ of the macroscopic spacetime structure $D_X$. Complementing this, $V_{max}$ is equivalent to the entourage of U generated by the entire set X. Thus, *a robust pregeometric correspondence between the microscopic spacetime structure and the macroscopic spacetime structure is provided.*

## 4. METRIC OVER Z2 x …x Z2

A nexus to conventional spacetime physics may obtain per the metric induced by $D_X$. As an example of this procedure, we generate the metric from a particular sequence $V_0$, $V_1$, ... constructed from elements of $U_B$ over the topological group Z2 x …x Z2. We chose this group structure, since the elements of $U_B$ are symmetric and thus elements of $D_X$ as required by Engelking's theorem. [Otherwise, per our pregeometric model of non-separability, it is necessary and sufficient that the entourages used to produce a metric simply conflate each $A_x$ with $A_x^{-1}$.]

The process is inductive, so we begin with Z2 x Z2, labeling its elements as follows:

$$e = (0,0)$$

$$a = (0,1)$$

$$b = (1,0)$$

$$c = (1,1).$$

Since Z2 x Z2 is a subgroup of Z2 x Z2 x Z2, we choose the following labels for Z2 x Z2 x Z2:

$$e = (0,0,0)$$

$$a = (0,0,1)$$

$$b = (0,1,0)$$

$$c = (0,1,1)$$

$$d = (1,0,0)$$

$$f = (1,0,1)$$

$$g = (1,1,0)$$

$$h = (1,1,1).$$

Next, we construct a sequence of entourages which satisfy $V_0 = X \text{ x } X$ and $(V_{i+1})^3 \subset V_i$. We choose the following sequence:

$$V_4 = A_a$$

$$V_3 = A_a \cup A_b$$

$$V_2 = A_a \cup A_b \cup A_c$$

$$V_1 = A_a \cup A_b \cup A_c \cup A_d$$

$$V_0 = A_a \cup A_b \cup A_c \cup A_d \cup A_f \cup A_g \cup A_h.$$

We now explicate the arguments producing a metric over Z2 x Z2 x Z2 which easily generalize to Z2 x …x Z2.

First, we are only concerned with members of $V_i$ which do not appear in $V_k$ with $k > i$, since we always seek to minimize the 'artificial' distance between adjacent members in a sequence of elements between x and y. Ignoring the diagonal we have:

(e, a), (b, c), (d, f), (g, h) $\in V_4$, i.e., all pairs producing a in the algebra of Z2 x Z2 x Z2

(e, b), (a, c), (d, g), (f, h) $\in V_3$, i.e., all pairs producing b       "               "

(e, c), (a, b), (d, h), (f, g) $\in V_2$, i.e., all pairs producing c       "               "

(e, d), (a, f), (b, g), (c, h) $\in V_1$, i.e., all pairs producing d       "               "

(e, f), (e, g), (e, h), … $\in V_0$, i.e., all pairs producing f, g, and h "                ".

Now, the distance between any of the elements in the (above) pairs of $V_4$ is $\frac{1}{16} = (½)^4$, and the sequence producing this distance is simply the end points. The distance between elements in the pairs of $V_3$ is $\frac{2}{16} = (½)^3$, also produced by the end points. The distance between elements in the pairs of $V_2$ can be found in sequences constructed from a pair in $V_4$ plus a pair in $V_3$, i.e., $\frac{1}{16} + \frac{2}{16} = \frac{3}{16}$. Note, this analysis holds for Z2 x Z2 with the entourage sequence $V_2 = A_a$, $V_1 = A_a \cup A_b$, and $V_0 = A_a \cup A_b \cup A_c$. One simply ignores elements d – h and the distances $\rightarrow$ $\frac{1}{4}, \frac{2}{4}$, and $\frac{3}{4}$. Note also there is no mixing between elements e – c and d – h in the these three entourages.

The first mixing we see between these subsets of elements occurs in $V_1$, i.e., those pairs producing d in the group structure. Since there is no mixing between these elements in smaller $V_i$ (larger i), any sequence between elements of the pairs in $V_1$ must contain at least one element of $V_1$. Thus, the distance between elements of these pairs is produced by a sequence which is the pair itself, i.e., the distance is $\frac{8}{16} = (\frac{1}{2})^1$. The pairs producing f, g, and h in the group structure are all that remain for evaluation and they are 'mixed' collections as well.

Thus, the sequence producing a distance between any of these pairs must contain at least one element of $V_1$, so the distance is greater than ($\frac{1}{2}$). If one views these pairs as an element from e – c and an element of d – h, then we should start a sequence in $V_i$ (i = 2,3,4) so as to have it end judiciously in $V_1$. We need, for example, the distance between e and f. We see that af = d, so we want a sequence beginning with e which gets us in minimal fashion to a, whence we can complete the sequence with (a, f) ∈ $V_1$. We have (a, e) ∈ $V_4$ which serves perfectly, producing a distance of $\frac{9}{16} = \frac{8}{16} + \frac{1}{16}$. This generalizes for all pairs producing f, since for xy = f and αy = d, we must have xα = a.

Similarly, we find for xy = g and αy = d that xα = b. This gives a distance between x and y of $\frac{10}{16} = \frac{8}{16} + \frac{2}{16}$. Finally, we find for xy = h and αy = d that xα = c. This gives a distance between x and y of $\frac{11}{16} = \frac{8}{16} + \frac{3}{16}$. To summarize:

$$g(x, y) = \frac{1}{16} \text{ for xy = a}$$

$$g(x, y) = \frac{2}{16} \text{ for xy = b}$$

$$g(x, y) = \frac{3}{16} \text{ for xy = c}$$

$$g(x, y) = \frac{8}{16} \text{ for xy = d}$$

$$g(x, y) = \frac{9}{16} \text{ for xy = f}$$

$$g(x, y) = \frac{10}{16} \text{ for } xy = g$$

$$g(x, y) = \frac{11}{16} \text{ for } xy = h$$

so our metric is in concert with the group structure. While tedious, this explication generates immediately a metric over Z2 x…x Z2 by induction. We have, for example, the following distances in the metric for Z2 x Z2 x Z2 x Z2 - $\frac{1}{64}$, $\frac{2}{64}$, $\frac{3}{64}$, $\frac{8}{64}$, $\frac{9}{64}$, $\frac{10}{64}$, $\frac{11}{64}$, $\frac{32}{64}$, $\frac{33}{64}$, $\frac{34}{64}$, $\frac{35}{64}$, $\frac{36}{64}$, $\frac{37}{64}$, $\frac{38}{64}$, and $\frac{39}{64}$. Notice that in general, the largest distance

$$g_{max} = \frac{2^{2k-2} + 2^k - 2}{2^{2k-1}}$$

where $N = 2^k$. Thus, $g_{max} \to \frac{1}{2}$ as $k \to 4$.

## 5. CONCLUSION

The uniform spaces $D_X$ and $U$ may be produced over any denumerable set X of spacetime events. These structures provide independent, but self-consistent, pregeometric definitions of macroscopic and microscopic spacetime neighborhoods respectively. The macroscopic spacetime neighborhoods serve as a pregeometric basis for classical spacetime physics, and the microscopic spacetime neighborhoods serve as a pregeometric basis for quantum physics. While heuristic, a transition from this pregeometry to conventional spacetime physics is perhaps intimated per the metric induced by $D_X$. Obviously, the process by which this metric yields a spacetime metric with Lorentz signature must be obtained.

While yet inchoate, this formalism implies the potential of pregeometry for elucidating QNS specifically and therefore, quantum gravity in general. Responding to Sorkin's "laundry list of alternatives concerning Quantum Gravity" (Sorkin, 1997) we have: (1) The deep structure of spacetime is discrete and non-local; (2) The most basic features of spacetime are described via uniformity, topology, and group structure; and (3) Spacetime topology is not dynamic. Items (1) and (2) are more strongly supported by the mathematics than is item (3).

·In support of (1) – the uniform space which yields a pregeometry consistent with QNS and guarantees a metric structure is generated over a denumerable, finite collection of spacetime events X. [Note that in this formalism, quantum non-separability is actually fundamental to quantum non-locality *a la* Healey (1989).]

·In support of (2) - the discrete uniformity over X provides a natural, non-metric notion of open balls and induces the topology for U. The choice of a group structure then establishes the degree to which the spacetime harbors QNS. [Algebraic structure is also the source of quantum non-locality in Heller's quantum gravity program (Heller and Sasin, 1999).]

Concerning item (3), the combinatorial nature of the pseudometric induced by $D_X$ is strongly suggestive of a stochastic pregeometry. This supports a dynamic topological structure for spacetime. However,

·In defense of (3) - choosing a combinatorial description of spacetime at the pregeometric level presupposes an 'ontological' basis for the stochasticity of QM *a la* the Copenhagen interpretation. Our formalism allows for the possibility that quantum gravity provides deterministic elements at the basis of QM *a la* Bohm's interpretation.

Finally, contrary to Bergliaffa *et al.* (1998), our pregeometric model of QNS suggests trans-temporal objects are not fundamental constituents of reality. In this sense, we agree with Gomatam (1999) who argues for a revision of our notion of macroscopic objects in accord with quantum non-separability. Indeed, the key to progress in quantum gravity may lie in a willingness to abandon stalwart concepts of dynamism such as energy, momentum, force, and even causation at the fundamental level of modeling.